# Efficient Photon Upconversion Enabled by Strong Coupling Between Organic Molecules and Quantum Dots

Kefu Wang[1#], R. Peyton Cline[2#], Joseph Schwan[3#], Jacob M. Strain[4], Sean T. Roberts[4*], Lorenzo Mangolini[3*], Joel D. Eaves[2*], & Ming Lee Tang[1*]

[1]*Department of Chemistry, University of Utah, Salt Lake City, USA; University of California, Riverside, Riverside, USA.*

[2]*Department of Chemistry, University of Colorado Boulder, Boulder, USA.*

[3]*Department of Mechanical Engineering, University of California Riverside, Riverside, USA.*

[4]*Department of Chemistry, The University of Texas at Austin, Austin, USA*

[#]*These authors contributed equally*

[*]*E-mail address: roberts@cm.utexas.edu; lmangolini@engr.ucr.edu; Joel.Eaves@colorado.edu; minglee.tang@utah.edu*

**Abstract**

Hybrid structures formed between organic molecules and inorganic quantum dots can accomplish unique photophysical transformations by taking advantage of their disparate properties. The electronic coupling between these materials is typically weak, leading photoexcited charge carriers to spatially localize to a dot or a molecule at its surface. However, we show that by converting a chemical linker that covalently binds anthracene molecules to silicon quantum dots from a carbon-carbon single bond to a double bond, we access a strong-coupling regime where excited carriers spatially delocalize across both anthracene and silicon. By pushing the system to delocalize, we design a photon upconversion system with a higher efficiency (17.2%) and lower threshold intensity (0.5 W/cm$^2$) than that of a corresponding weakly-coupled system. Our results show that strong coupling between molecules and nanostructures achieved through targeted linking chemistry provides a new route for tailoring properties in materials for light-driven applications.

In recent years, researchers have functionalized quantum dots (QDs) with molecules to generate new hybrid materials[1–5] for applications in solar energy harvesting,[6–9] catalysis,[10–14] and light emission.[15–17] These structures combine advantageous electronic properties of QDs, which possess high absorption cross-sections and size-tunable optical gaps, with those of molecules, which can exhibit high energy transfer efficiencies and specific chemical reactivities. In hybrid structures, the transfer of energy or charge transfer between QDs and molecules at their surfaces is often critical to their function. Typically, molecules are adhered to QDs via non-covalent van der Waals or ionic interactions. Such weak bonding implies weak electronic coupling between them. In the weak-coupling regime, the wavefunctions of excited charge carriers are spatially localized to either the QD or surface-anchored molecules. Because electronic coherences are short-lived in the weak-coupling limit, energy or charge moves between QDs and molecules via discrete, incoherent hops, as described in theories developed by Marcus,[18] Förster,[19] and Dexter.[20]

However, if the electronic coupling between a QD and molecules can be amplified, fundamentally different electronic states can emerge. Due to hybridization of their electronic wavefunctions, charge carriers are simultaneously shared between the molecule and QD. In this strong-coupling regime, a QD and surface-anchored molecules do not behave as separate entities, but rather as single material whose electronic properties are distinct from those of its individual components. In the solid state, this scenario is analogous to the formation of an alloy, where strong electronic coupling between distinct atoms yields a new material with unique functionality.

In this manuscript, we demonstrate that by controlling the structure of carbon bridges that anchor anthracene molecules to silicon QDs, we produce strongly-coupled triplet excitons – spin-1 electron-hole bound states – that spatially delocalize across both materials. Triplet states in QD:molecule hybrid systems hold strong interest due to their utility in photon upconversion systems that covert red to near-infrared light into UV-to-visible emission.[2,3,11,16,17,21,22] In these systems, QD:molecule hybrids function as sensitizers that absorb long-wavelength photons and pass their energy in the form of triplet excitons to emitter molecules in solution. Pairs of these excited emitters subsequently pool their energy to produce short-wavelength emission via triplet fusion.

By using a π-conjugated carbon bridge to link anthracene molecules to silicon QDs, we create a strongly-coupled hybrid. This enhanced coupling impacts both the energy and spatial distribution of spin-triplet excitons formed by this system; as evidenced by measurable effects in steady-state and time-resolved optical experiments, and in electronic structure calculations that all differ qualitatively from those of systems lacking strong coupling. By varying the energy of strongly-coupled triplet excitons via altering the number of anthracene molecules that couple to silicon QDs, we design a photon upconversion system that converts green light to blue, achieving an efficiency (17.2%) and threshold power (0.5 W/cm$^2$) that surpass values obtained for prior silicon QD-based systems.[23]

Our results represent the first example of a QD:molecule system that exhibits strongly-coupled excitonic states with well-defined triplet character. For hybrid QD:molecule systems, strong coupling can be advantageous because it allows for both tunability over their electronic properties and, through coherence, can bypass metastable intermediates in energy or charge transfer that lead to loss pathways. We anticipate the silicon QD:anthracene system we report, which uses extended π-conjugation between silicon and carbon to achieve exciton delocalization, can be used to produce strongly-coupled nanoscale objects with designer electronic properties for applications in energy conversion, optoelectronics, and catalysis.

Silicon QDs were selected for this work as we hypothesized their ability to form strong,

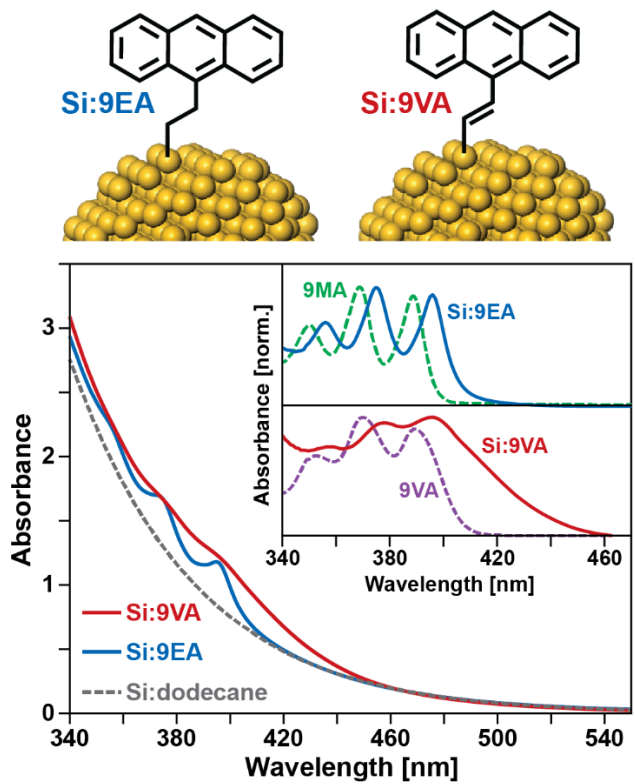

**Figure 1**: (Top) Structures of Si:9EA and Si:9VA. (Bottom) Absorption spectra of Si:dodecane (grey dashed), Si:9EA (blue, $\langle N_{9EA} \rangle$ = 3.0), and Si:9VA (red, $\langle N_{9VA} \rangle$ = 5.3) in toluene. (Inset) Absorption spectra of 9-methylanthracene (9MA) and 9-vinylanthracene (9VA) after subtracting the Si:dodecane background. 9MA and 9VA serve as molecular references for surface-anchored anthracene molecules in Si:9EA and Si:9VA respectively. The broadening of the resonances in Si:9VA are one indication of strong coupling.

covalent bonds with carbon where electrons are equally shared[24,25] would allow for increased electronic coupling between them and molecules anchored to their surface. This ability sets silicon apart from other common QD materials, such as CdSe and PbS, which coordinate molecules to their surface via ionic interactions wherein electrons are localized on one side of the QD:molecule bridge.[3] Silicon QDs used in this work were prepared via a non-thermal plasma synthesis and functionalized in-flight with 1-dodecene to yield dodecane-capped silicon QDs (Si:dodecane) that were soluble in hydrophobic solvents.[26,27] Si:dodecane was subsequently hydrosilylated with 9-ethynylanthracene, yielding silicon QDs functionalized with a mixture of aliphatic dodecane and 9-vinylanthracene ligands. We refer to these surface-functionalized QDs as Si:9VA (**Figure 1**).

As a control, a second set of silicon QDs was prepared by replacing 9-ethynylanthracene with 9-vinylanthracene, which yielded silicon QDs with anthracene ligands attached by a 2-carbon chain whose carbon atoms are $sp^3$ hybridized rather than $sp^2$. We refer to this control sample as Si:9EA (**Figure 1**).

**Figure 1** shows absorption spectra of Si:dodecane, Si:9EA, and Si:9VA in toluene. The absorption spectrum of Si-dodecane is relatively featureless (**Figure 1**, grey dashed), reflecting the indirect bandgap of silicon. Examining spectra of Si:9EA (**Figure 1**, blue), a series of resonances appear at 396, 375, and 356 nm that correspond to the vibrational fine structure of the $S_0 \rightarrow S_1$ transition of 9-ethylanthracene (9EA) molecules bound to silicon. As noted in prior work,[23] these resonances are slightly red-shifted relative to corresponding features in the absorption spectrum of 9-methylanthracene (**Figure 1**, inset), but otherwise do not differ in their linewidth or oscillator strength, indicating that electronic coupling between 9EA molecules and the silicon QDs to which they are bound is weak.

In contrast to 9EA, 9-vinylanthracene (9VA) molecules display distinctly different behavior. Upon binding to silicon, each of the absorption peaks that form the vibrational substructure of 9VA's $S_0 \rightarrow S_1$ transition show significant broadening (**Figure 1**, red). The extended π-conjugation of Si:9VA's $sp^2$ linkage is expected to facilitate spatial overlap of states involving 9VA's π-electrons and those of silicon, improving their electronic coupling. If sufficiently strong, this coupling can induce partial hybridization of the valence states of 9VA with several silicon states, leading each isolated 9VA absorption resonance to broaden into a band of states with mixed silicon:anthracene character. Thus, the spectral broadening we observe is suggestive of improved silicon:anthracene coupling in Si:9VA with respect to Si:9EA (**Figure 1**, inset).

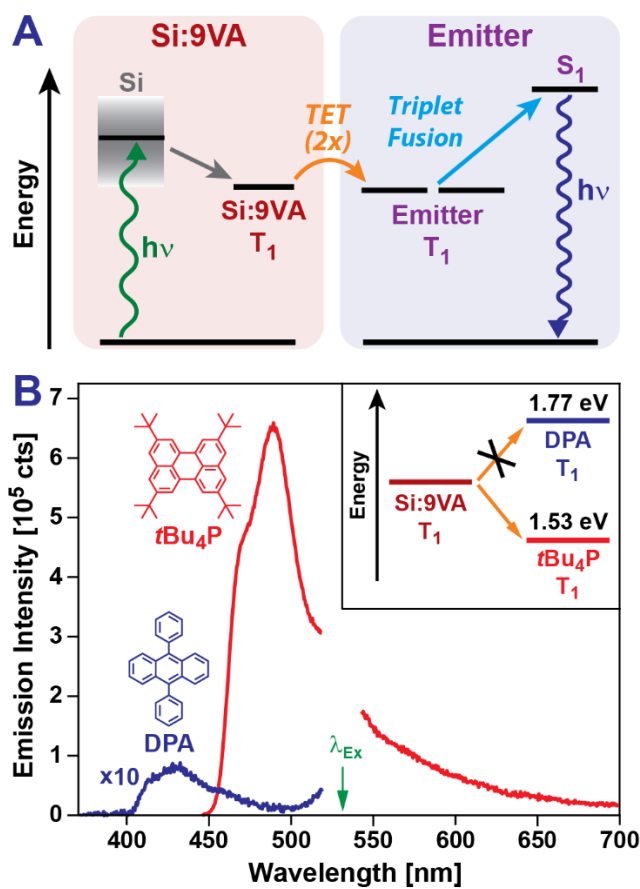

**Figure 2:** (A) Energy level diagram illustrating photon upconversion in Si:9VA. Photons absorbed by Si:9VA fuel the transfer of triplet excitons to emitter molecules that subsequently undergo triplet fusion to produce emissive, high-energy, spin-singlet states. (B) Emission spectra ($\lambda_{Ex}$ = 532 nm) of mixtures of Si:9VA ($\langle N_{9VA} \rangle$ = 7.3) with DPA (5.2 mM, blue) and $t$Bu$_4$P (5.2 mM, red). While negligible upconversion is obtained from DPA, efficient upconversion is seen from $t$Bu$_4$P. (Inset) Energy level diagram highlighting that a decrease in Si:9VA's triplet exciton energy due to strong coupling will hinder energy transfer to DPA while still allowing energy transfer to $t$Bu$_4$P.

We hypothesized that the stronger coupling present in Si:9VA would improve its ability to shuttle energy between its silicon QD core and anthracene molecules at its surface, which in turn should lead to its superior performance in triplet-fusion based photon upconversion systems (**Figure 2A**). In these systems, photons absorbed by Si:9VA are used to produce excitons with spin-triplet character that can be subsequently passed to emitter molecules diffusing in solution. Diffusive encounters between pairs of excited emitters can allow them to undergo triplet fusion, producing a high-energy, luminescent spin-singlet state.

**Figure 2B** plots emission spectra obtained from toluene solutions containing Si:9VA and one of two known emitters, 9,10-diphenylanthacene (DPA) or 2,5,8,11-tetra-tert-butylperylene ($t$Bu$_4$P). Previously, we showed Si:9EA can efficiently drive triplet exciton transfer to DPA, fueling production of upconverted light with 7% efficiency.[23] We note the in-flight functionalized Si QDs we employ for this work have allowed us to improve that yield to 15.8%. In contrast, we find Si:9VA particles achieve a paltry upconversion quantum efficiency of only 0.03% when paired with DPA (**Figure 2B**, blue). This result appears to go against our expectation that stronger coupling between anthracene and silicon in Si:9VA should lead to improved energy transfer.

As photon upconversion involves multiple energy transfer steps, to identify which step was responsible for the reduced performance of Si:9VA, we replaced DPA in our upconversion system with $t$Bu$_4$P. Doing so produces an upconversion yield, 2.7%, that is over two orders of magnitude larger than that observed when paring Si:9VA with DPA. This result suggests it is not energy transfer from silicon to 9VA that limits the DPA upconversion system, but rather energy transfer from Si:9VA to DPA. Importantly, DPA and $t$Bu$_4$P possess different triplet energies of 1.77 eV[28] and 1.53 eV,[28–32] respectively. While anthracene has a triplet energy of 1.8 eV[28,33] that is just high enough in energy to sensitize triplet energy transfer to DPA, if this energy is lowered in Si:9VA due to coupling to silicon, it will hinder its ability to transfer energy to DPA while having a smaller impact on energy transfer to $t$Bu$_4$P.

To evaluate if the sp$^2$ bridge of Si:9VA lowers its triplet exciton energy below that of Si:9EA, we examined their electronic structures using density functional theory (DFT). Our computational model balances the demands of simulating a large system against the extraordinary computational cost of computing electronically excited states accurately. It consists of a single 9VA or 9EA molecule attached to a semiperiodic Si(111) surface (**Figure 3**). We chose to employ this surface as it corresponds to silicon's lowest energy surface facet.[34–37] To mimic a silicon QD, we use a slab comprised of 6 crystalline layers, periodically replicated in two dimensions. As we are interested in triplet exciton states, we compute the solutions of the Kohn-Sham equations for the triplets in each Si(111):molecule system using spin-polarized DFT and compute the approximate exciton density of states from the conduction and valence bands.[38–40]

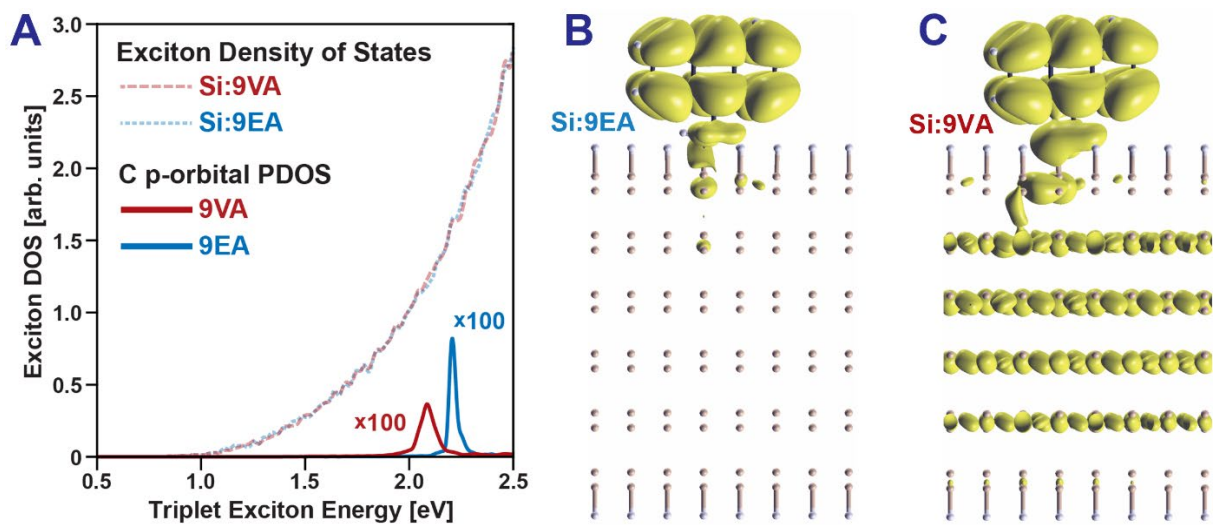

**Figure 3:** (A) Density of triplet exciton states computed for Si:9VA (light red dashed) and Si:9EA (light blue dotted). Contributions from the carbon p-orbitals come from atom-projected density of states for 9VA (solid red) and 9EA (solid blue). These projections have contributions from the principal frontier orbitals of the anthracene and linker (carbon atom p-orbital, projected density of states, PDOS). As the sharp resonances of the molecular states mix with those of the solid, the PDOS shifts and broadens. (B & C) Band-decomposed "partial" charge densities that highlight the contribution of holes to excitonic states that contain significant carbon p-orbital character for (B) Si:9EA and (C) Si:9EA. They display weak mixing between anthracene and silicon states in Si:9EA and strong coupling in Si:9VA. Weak coupling in Si:9EA leads to states that remain localized to anthracene while strong coupling in Si:9VA gives rise to delocalized states that span the organic:inorganic interface.

**Figure 3A** displays the density of triplet exciton states (DOS) computed for both Si(111):9EA and Si(111):9VA. To identify states involving surface-bound anthracene molecules, we have projected out contributions to each computed state from the p-orbitals of the carbon atoms of 9EA and 9VA. In the absence of electronic coupling between silicon and anthracene, the DOS involving these orbitals will be sparse and sharply peaked, just as they are in individual anthracene molecules where large energy gaps separate each of their triplet states. However, if anthracene couples to silicon, the localized states of the molecule will hybridize with several states of the silicon, spreading spatially and energetically, creating a new band of states with mixed silicon:anthracene character. The greater the shifting and broadening of the molecular peaks is, the greater the strength of the coupling between the molecule and silicon.

By projecting the density of states (PDOS) onto the carbon p-orbitals, we find for Si:9EA (**Figure 3A, blue**) that they fall within a narrow range in energy, indicating that 9EA only weakly couples to silicon. In comparison, the PDOS broadens significantly for Si:9VA, indicating stronger coupling between anthracene and silicon in this system (**Figure 3A, red**). As expected, this coupling shifts the energy of triplet states involving anthracene in Si:9VA, lowering it below the energy of these states in Si:9EA. This is in accordance with our photon upconversion results, which suggest the triplet exciton energy of Si:9VA is lower than that of Si:9EA (**Figure 2B**).

To further explore how silicon:anthracene electronic coupling in Si:9VA and Si:9EA impacts the spatial distribution of their excitonic states, we have computed the band-decomposed "partial" charge densities for triplet states containing carbon p-orbital character. In **Figures 3B & 3C**, we plot the contribution of holes to these states for Si:9EA and Si:9VA using the XCrySDen software package[41] (see supporting information, section 4 for computational details). While holes remain well localized to 9EA (**Figure 3B**), we see significant extension of charge density across the Si:9VA interface (**Figure 3C**). This indicates the electronic coupling in Si:9VA is sufficiently large to push the system into the strong-coupling limit, giving rise to triplet exciton states of mixed silicon:anthracene character that spatially delocalize across these two materials.

Experimental results from transient absorption (TA) spectroscopy in Si:9VA are consistent with the qualitative assignments of strong coupling from DFT. Previously, we showed that photoexcitation of silicon QDs functionalized with 9EA produces a QD-centered triplet exciton state that transfers from silicon to 9EA on a 15.2 ns timescale.[23] This transfer occurs in the weak coupling limit as evidenced by the appearance of an induced absorption band that agrees well with the triplet exciton absorption spectrum of 9-methylanthracene (**Figure 4A, left** & **Figure 4B, blue**).

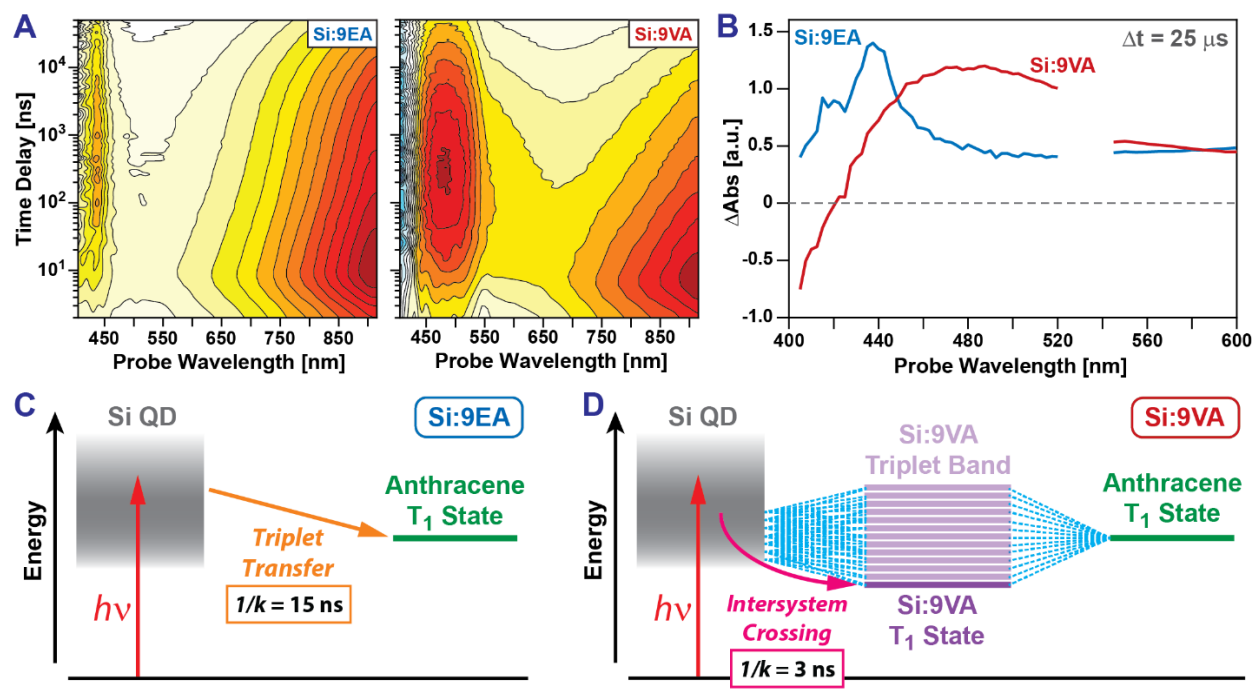

**Figure 4**: (A) TA spectra measured following 532 nm photoexcitation of (left panel) Si:9EA ($\langle N_{9EA} \rangle$ = 3.0) and (right panel) Si:9VA ($\langle N_{9VA} \rangle$ = 5.3). (B) TA spectra recorded at a time delay of 25 μs of Si:9EA (blue) and Si:9VA (red). These spectra highlight the nature of the triplet exciton state formed by each system. In Si:9EA, a vibronic progression associated with 9EA's triplet state appears at 438 and 418 nm and signals weak coupling between silicon and 9EA. In Si:9VA, a vibronic progression is not observed. Rather, a broad featureless peak indicates strong electronic coupling between silicon and 9VA. (C & D) Energy level diagrams illustrating the behavior of (C) Si:9EA and (D) Si:9VA following photoexcitation. In Si:9EA, coupling between anthracene and silicon is weak. Photoexcitation of silicon leads to population of a triplet state localized on 9EA on a 15 ns timescale. In Si:9VA, strong coupling between anthracene and silicon gives rise to a band of triplet states with mixed silicon:anthracene character. Photoexcitation of silicon populates the lowest energy state of this band on a 3 ns timescale via intersystem crossing.

In contrast, we observe fundamentally different behavior for Si:9VA (**Figure 4A, right**). Photoexciting Si:9VA at 532 nm generates charge carriers within the silicon QD, producing a broad induced absorption peaked near 950 nm that stems from interband transitions of these carriers. Over a 3 ns timescale, this band decays and a new induced absorption feature centered near 480 nm appears along with a photobleach at wavelengths shorter than 420 nm (**Figure 4B, red**). This bleach agrees well with the ground state absorption onset of 9VA molecules that are bonded to silicon (**Figure 1, inset**), indicating the induced absorption band at 470 nm arises from an excited state in which these molecules participate. Notably, this band's spectral lineshape is distinct from that of a triplet excitation localized on 9VA (**Figure S3**) and from a charge transfer state wherein an electron or hole has been donated from silicon to 9VA.[42,43] This suggests the generated excited state differs from one wherein charge carriers have been fully transferred from silicon to 9VA. Rather, this band is reminiscent of Fano-type lineshapes that result when the energetically sparse electronic states of a molecule mix with the dense manifold of electronic states of a semiconductor or metal,[44,45] signifying strong coupling between 9VA and silicon. This indicates the 3 ns timescale we observe for formation of this state arises not from energy transfer from silicon to 9VA but rather represents the timescale for intersystem crossing that converts a spin-singlet state localized on the silicon QD to a spin-triplet state that spatially extends across the silicon QD:9VA interface (**Figure 4D**).

Together, our electronic structure calculations and time-resolved measurements indicate that Si:9VA's sp$^2$ linkage engenders strong coupling between silicon QDs and anthracene and that this coupling acts to lower the triplet exciton energy of this system. With this knowledge in hand, we can fine tune this system's electronic structure to optimize it for photon upconversion. As each 9VA molecule that binds to a silicon QD couples to it strongly, the band of strongly-coupled triplet states will widen and shift to lower energy as the number of surface-bound 9VA molecules is increased. Hence, by varying the number of bound 9VA molecules, we expect we can control the energetic position of Si:9VA's lowest-energy triplet exciton state ($T_1$) and hence set the energetic driving force for triplet exciton transfer from Si:9VA to an upconversion emitter (**Figure 5A**).

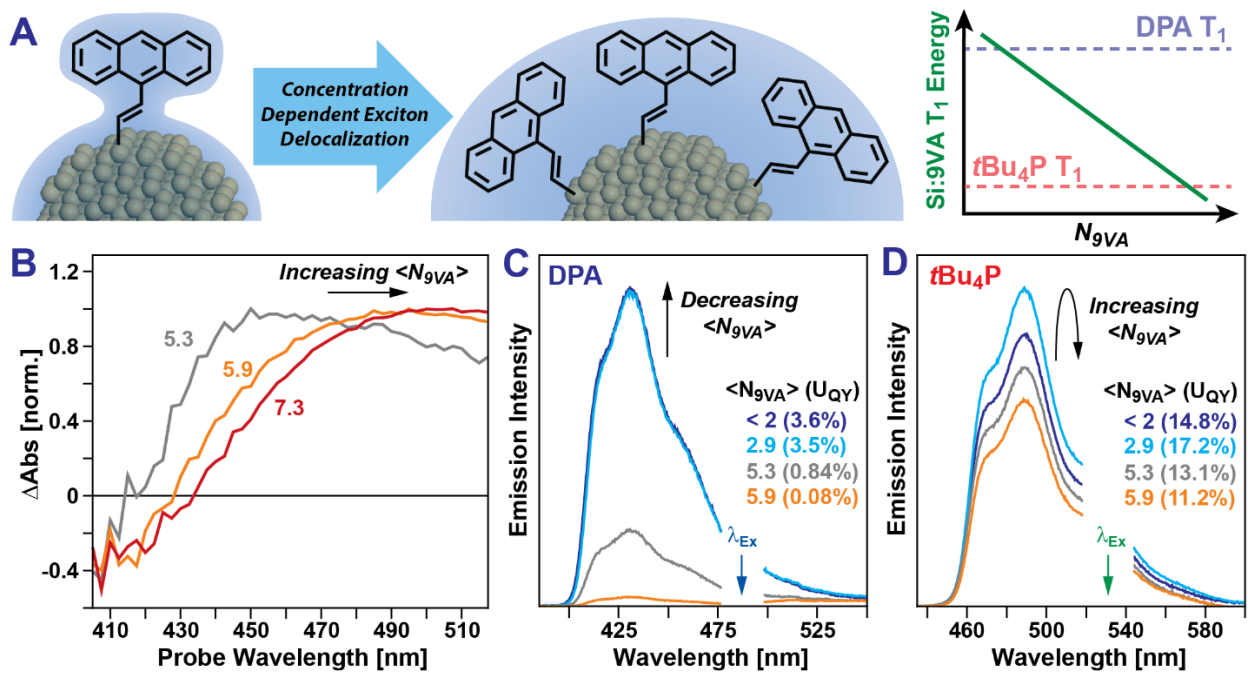

**Figure 5:** (A) By increasing the number of strongly-coupled 9VA molecules that bind to a silicon QD, we can tune the energy of Si:9VA's $T_1$ state, bringing it into energetic alignment with triplet acceptors. (B) TA spectra of Si:9VA at a time delay of 25 μs measured as a function of 9VA surface concentration ($\langle N_{9VA} \rangle$ = 5.3, 5.9, 7.3). A red-shift of the induced absorption of the Si:9VA triplet exciton band with increasing $\langle N_{9VA} \rangle$ indicates that Si:9VA's $T_1$ state energy is dependent on the number of 9VA molecules bound to silicon. (C & D) Upconversion emission spectra of DPA (5.2 mM, $\lambda_{Ex}$ = 485 nm) and $t$Bu$_4$P (5.2 mM, $\lambda_{Ex}$ = 532 nm) sensitized by Si:9VA particles with differing 9VA surface concentrations. By lowering $\langle N_{9VA} \rangle$ we improve upconversion emission by DPA by raising Si:9VA's $T_1$ state energy. By raising $\langle N_{9VA} \rangle$, we improve upconversion emission by $t$Bu$_4$P up to a point by lowering Si:9VA's $T_1$ energy.

To test this concept, we have measured TA spectra of silicon QDs functionalized with different surface concentrations of 9VA (**Figure 5B**). As the average number of surface-bound 9VA molecules, $\langle N_{9VA} \rangle$, is increased from 5.3 to 7.3, we observe a progressive redshift of the photoinduced absorption stemming from Si:9VA's strongly-coupled triplet exciton state, from 470 nm to 508 nm. Such a shift is unexpected if 9VA surface concentration has no impact on Si:9VA's electronic structure. Indeed, for Si:9EA particles wherein anthracene molecules only weakly couple to silicon, we find there is no dependence of the spectral position of the 9EA triplet exciton absorption on $\langle N_{9EA} \rangle$ (**Figure S5**). In contrast, the spectral shift observed in Si:9VA can be explained if increased 9VA surface concentration alters Si:9VA's triplet exciton band structure

(**Figure 5B**).

Spurred on by our TA results, we have optimized the performance of Si:9VA-based photon upconversion systems by varying the energy of Si:9VA's T$_1$ state energy by controlling $\langle N_{9VA} \rangle$. **Figures 5C** and **5D** respectively highlight the performance of upconversion systems that employ DPA and *t*Bu$_4$P emitters. Whereas Si:9VA particles that bind on average 7 9VA molecules exhibited a miniscule upconversion efficiency of 0.03% when paired with DPA (**Figure 2B**), we find that by reducing $\langle N_{9VA} \rangle$ to less than 2 raises the upconversion efficiency by over two orders of magnitude, to 3.6% (**Figure 5C**). We attribute this gain to a reduction in the broadening of Si:9VA's triplet exciton band structure, which raises the energy of Si:9VA's T$_1$ state and allows it to better donate energy to DPA.

In contrast to behavior seen for DPA, we find increasing $\langle N_{9VA} \rangle$ from < 2 to 3 enhances photon upconversion by *t*Bu$_4$P (**Figure 5D**), which achieves a yield of 17.2% after accounting for inner filter effects (**Figure S1**). We note this yield surpasses that of the most efficient silicon QD-based upconversion system known prior to this work, Si:9EA particles paired with DPA.[23] The performance of Si:9VA paired with *t*Bu$_4$P is even more impressive given that DPA has a higher fluorescence quantum yield than *t*Bu$_4$P (95% vs. 70%),[23,46] which means that all things being equal, we would expect DPA-based systems to be roughly a third more efficient at producing upconverted emission than *t*Bu$_4$P-based systems.

Interestingly, increasing $\langle N_{9EA} \rangle$ beyond 3 leads to a moderate drop in upconversion produced by *t*Bu$_4$P. This decrease may stem from a widening of Si:9VA's triplet exciton band structure to a point wherein triplet energy transfer to *t*Bu$_4$P becomes thermally activated or could also result from formation of low-energy aggregate states between 9VA molecules that directly interact with one another on the particle's surface. Aggregate formation has been reported on the surfaces of $SiO_2$ particles[47] and implicated to play a role in both triplet transfer[48] and electron transfer[49] between molecules and semiconductors, but assessing the involvement of such states in the photoexcited dynamics of Si:9VA extends beyond the scope of this report.

In summary, by controlling the nature of the chemical bond that affixes anthracene molecules to silicon QDs, we have achieved strong electronic coupling between these materials. Strong coupling enables formation of triplet exciton states that spatially extend across the silicon:anthracene interface. The energy of these spatially-delocalized triplet exciton states can be altered by varying the number of strongly-coupled anthracene molecules that bind to silicon. By controlling this energy, we optimize their ability to fuel photon upconversion, achieving a record efficiency yield for silicon QD based upconversion systems (17.2%) at a power threshold of only 0.5 W/cm$^2$. More broadly, strong coupling between QDs and molecules provides a complementary handle for tuning the electronic structure of nanomaterials, in addition to changing their size and shape. By delocalizing charge carriers across all molecules bound to a QD, strong coupling can provide unique opportunities for creating systems with an enhanced ability to donate and accept charge, drive chemical transformations, and reshape the energy content of light.

**Authorship Statement**

K.W. conducted the nanocrystal functionalization, photon upconversion, and ns-transient absorption (TA) experiments. R.P.C. did the calculations; J.S. the non-thermal plasma synthesis; J.M.S. the sub ns-TA. K.W., J.S., L.M. and M.L.T. conceived of the project. R.P.C. and J.D.E. designed the calculations. S.T.R. composed the manuscript with significant contributions from all authors.

**Data Availability:** Experimental and computed data as well as MATLAB software used to analyze and plot data are available from the authors upon request.

**Corresponding Authors**

Sean T. Roberts: OCRID ID, 0000-0002-3322-3687; email, roberts@cm.utexas.edu

Lorenzo Mangolini: OCRID ID, 0000-0002-0057-2450; email: lmangolini@engr.ucr.edu

Joel D. Eaves: OCRID ID, 0000-0002-9371-1703; email: Joel.Eaves@colorado.edu

Ming Lee Tang: OCRID ID, 0000-0002-7642-2598; email, minglee.tang@utah.edu

**Acknowledgements**

This work was supported under NSF grant CMMI-2053567. Work at the University of Texas at Austin was additionally supported by the Welch Foundation (Grant F-1885) and W. M. Keck Foundation (Grant 22605). Work at the University of California, Riverside was also supported by AFOSR grant FA9550-20-1-0112. This work utilized the Summit supercomputer, which is supported by the National Science Foundation (awards ACI-1532235 and ACI-1532236), the University of Colorado Boulder, and Colorado State University. The Summit supercomputer is a joint effort of the University of Colorado Boulder and Colorado State University.

# Efficient Photon Upconversion Enabled by Strong Coupling Between Organic Molecules and Quantum Dots

Kefu Wang[1#], R. Peyton Cline[2#], Joseph Schwan[3#], Jacob M. Strain[4], Sean T. Roberts[4*], Lorenzo Mangolini[3*], Joel D. Eaves[2*], Ming Lee Tang[1*]

[1]*Department of Chemistry, University of Utah, Salt Lake City, USA; University of California, Riverside, Riverside, USA.*

[2]*Department of Chemistry, University of Colorado Boulder, Boulder, USA.*

[3]*Department of Mechanical Engineering, University of California Riverside, Riverside, USA.*

[4]*Department of Chemistry, University of Texas at Austin, Austin, USA*

[#]*These authors contributed equally*

[*]*E-mail address: roberts@cm.utexas.edu; lmangolini@engr.ucr.edu; Joel.Eaves@colorado.edu; minglee.tang@utah.edu*

## 1. Instrumentation

Absorption spectra were recorded on a Cary 5000 UV-Vis absorption spectrophotometer. Photoluminescence (photon upconversion, power dependence) spectra were recorded on a Maya 2000-Pro Spectrometer (Ocean Optics Inc.) following excitation by continuous wave (CW) solid-state lasers (488 nm: OBIS LX 75 mW, Coherent Inc.; 532 nm: Sapphire SF 532, Coherent Inc.). Neutral density filters (Thorlabs) were used to tune the power of the excitation source without changing its beam size. Semrock notch filters were used to remove the excitation light following the sample. Laser power was measured using a benchtop optical power and energy meter (2936R, Newport Corp.) and silicon wand detector head (818-ST2/DB, Newport Corp.).

Nuclear magnetic resonance (NMR) spectra were recorded on a Bruker AVANCE NEO 400 MHz NMR spectrometer at room temperature. $^1$H chemical shifts (δ) are reported in parts per million with a residual solvent ($CDCl_3$) peak as an internal standard.

All transient absorption (TA) measurements with the exception of data highlighted in **Figure S5** were conducted using an enVISion spectrometer from Magnitude Instruments that employed a 532 nm excitation laser. The repetition rate of the excitation laser was varied from 1 kHz to 19 kHz depending on the size of the delay time window scanned, with lower repetition rates used for longer time windows. TA spectra highlighted in **Figures 4**, **5**, **S3**, and **S4** were recorded using a 10 kHz repetition rate and a pump fluence of 150 μJ/cm$^2$. The instrument response function for these measurements was found to give a time resolution of 4.2 ns. Higher time resolution TA experiments with an instrument response of ~1.6 ns (**Figure S5**) were performed using 532 nm pump pulses generated by a frequency-doubled Q-switched Nd:YAG laser (Alphalas Pulselas-A, < 1 ns, 8.7 μJ). Supercontinuum probe pulses were produced by focusing the output of a Ti:sapphire amplifier operating at 804 nm (Coherent Duo Legend Elite, 3 kHz, 4.5 mJ) into a flowing liquid cell filled with $D_2O$ (Starna). A digital delay generator (Stanford Research Systems DG535) was used to synchronize the operation of the pump and probe lasers and to vary their time of arrival at the sample. A silicon CCD (Princeton Instruments PyLoN 100-BR) interfaced with a 500 mm Czerny-Turner spectrometer (Acton Instruments Spectra Pro 2556) was used to detect pump-induced changes in the transmission of the probe through the sample. All silicon QD samples used for TA measurements were prepared in a nitrogen glove box and placed into sealed cuvettes.

## 2. Sample Preparation
### 2.1 Materials

9-vinyl anthracene (9VA, 97%) and 9,10-diphenylanthracene (DPA, 98%) were purchased from TCI America. Platinum octaethylporphyrin (PtOEP, 98%) used for triplet sensitization experiments and 2,2'-Azobis(2-methylpropionitrile) (AIBN) came from Sigma Aldrich. Deuterated chloroform ($CDCl_3$, 99.8%) was obtained from Cambridge Isotope Laboratories, Inc. Toluene was purchased from Fisher and dried and degassed with a JC Meyer solvent purification system. Dried methanol was procured from Sigma Aldrich and degassed before moving it into a glove box. Mesitylene (98%) was likewise obtained from Sigma Aldrich and was dried with molecular sieves. All other chemicals were used as received.

### 2.2 Synthesis of 9-acetylene anthracene (9AA)

9-acetylene anthracene (9AA) was synthesized from 9-bromoanthracene according to Nierth et al.[1] NMR spectra of synthesized 9AA were found to be consistent with this prior published report.

### 2.3 Surface Functionalization of Silicon Quantum Dots (Si QDs)

Si QDs partially functionalized with 1-dodecane (Si:dodecane) were synthesized using a non-thermal plasma reactor following the approached described in Refs. 2 and 3. In this process, silicon nanocrystals are first nucleated and grown in a low-temperature plasma using silane gas ($SiH_4$) as a precursor. Hydrogen gas saturated with 1-dodecene vapor is introduce in the plasma afterglow to graft alkyl chains to QDs. The flow rate of the saturated hydrogen gas was optimized to ensure that the QDs form a stable colloidal dispersion in mesitylene, all while having a sufficient fraction of their surface still available for additional modification.

After synthesis, these QDs were transferred to a glovebox under a nitrogen environment and dissolved in toluene. The resulting concentrated Si:dodecane solution was then diluted with mesitylene until its optical density at 488 nm was ~1.2 in a 1 cm pathlength cuvette, after which it was heated at 180 ºC for 1.5 hours to reduce Si QD surface defects. To prepare Si QDs with different anthracene surface concentrations, varying amounts of 9VA and 9AA were used during Si QD hydrosilylation (**Table S1**). We found that these in-flight functionalized Si QDs were amenable to thermal hydrosilylation at 180 ºC using 9VA, but not 9AA as uncontrolled ligand polymerization occurred. We addressed this by performing radical initiated hydrosilylation using 2,2'-azobis(2-methylpropionitrile) (AIBN) for both 9VA and 9AA at 60 ºC (**Scheme S1**). All precursors needed to functionalize Si QDs via radical initiated hydrosilylation were added to a 20 mL vial with a Teflon cap, which was then sealed and heated to 60 ºC overnight while undergoing constant stirring. This induces Si hydrosilylation with 9VA and 9AA, thus forming 9-ethyl anthracene (9EA) attached to Si QDs (Si QD-C-C-anthracene, labeled as Si:9EA) and 9-vinyl anthracene (9VA) attached to Si QDs (Si QD-C=C-anthracene, labeled as Si:9VA), respectively. After the reaction, methanol was added to the clear yellow mixture (methanol:mixture = 2:1 by volume) to precipitate out the QDs by centrifuging at 12,000 rpm for 30 mins. The precipitated QDs are then redispersed in 1.5 mL of toluene, followed by addition of 3 mL of methanol and centrifuged at 12,000 rpm for 20 mins again. This cleaning procedure was repeated three times and the final Si:9EA or Si:9VA samples were redispersed in toluene for future use.

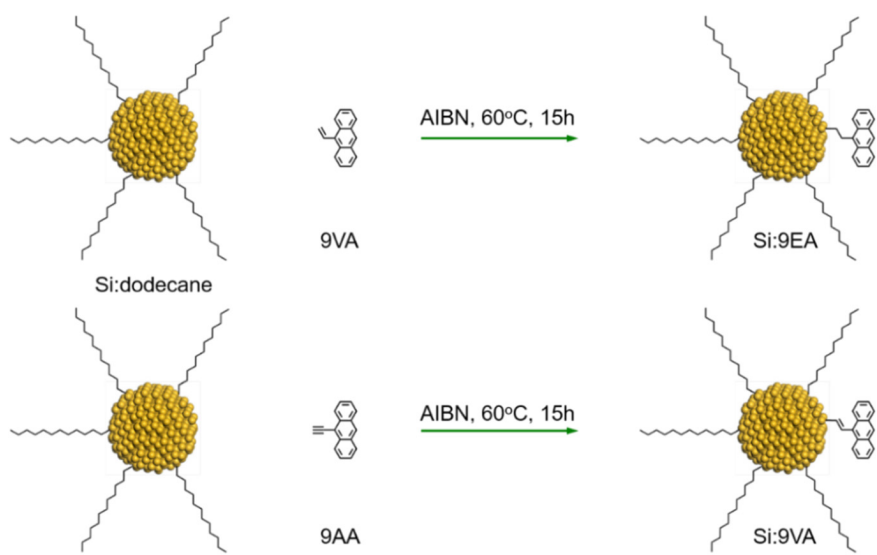

**Scheme S1:** Schematic illustration of Si QD hydrosilylation with 9VA and 9AA.

For photon upconversion measurements, Si:9EA solutions were diluted with toluene to give them an optical density of ~0.1 at 488 nm in a 1 cm pathlength cuvette. At that point, 9,10-diphenylanthacene (DPA) was added to the Si:9EA solution to give a final DPA concentration of 5.2 mM. The resulting mixture was transferred into a 10 mm × 10 mm quartz cuvette sealed with an air-tight Teflon cap in a glovebox before taking out for testing. The same procedure was applied to the preparation of Si:9VA with 2,5,8,11-tetra-tert-butylperylene (tBu$_4$P) emitters (5.2 mM).

|  | Si QDs (OD$_{488}$ = 1.2 ) | 9VA or 9AA (1 mg/mL) | AIBN (0.1 mg/mL) |
| --- | --- | --- | --- |
| Sample 1 | 200 µL | 8 µL (8 µg) | 6 µL |
| Sample 2 | 200 µL | 31 µ (31 µg) | 25 µL |
| Sample 3 | 200 µL | 125 µL (125 µg) | 100 µL |
| Sample 4 | 200 µL | 500 µL (500 µg) | 400 µL |

**Table S1:** Parameters used to prepare Si QDs with different surface ligand content via hydrosilylation in dry and degassed mesitylene. 2.000 mL of mesitylene was used for each reaction.

### 3. Characterization
### 3.1. Quantification of Number of Surface-bound Anthracene Molecules

The average number of anthracene molecules that bind to the surfaces of Si:9VA and Si:9EA was assessed using absorption spectroscopy. As highlighted in **Figure 1** of the main text, absorption spectra of Si:9VA and Si:9EA show distinct vibronic progressions in the range of 340 – 460 nm that arise from surface-bound anthracene molecules. The amplitude of these features can be obtained by subtracting the absorption spectrum of Si:dodecane from that of Si:9VA and Si:9EA samples. Prior to subtraction, to normalize for variations in the Si QD concentration of each sample, the Si:dodecane spectrum was scaled to match that of Si:9EA and Si:9VA in the spectral region between 550 – 700 nm wherein neither 9EA nor 9VA absorbs light. For Si:9EA, we can compare this background subtracted spectrum to reported molar extinction coefficients for 9-methylanthracene (9MA) to determine the concentration of surface-bound 9EA molecules in solution while the Si QD concentration can be determined using reported extinction spectra. This yields the following expression for $\langle N_{9EA} \rangle$, the average number of surface-bound 9EA molecules:

$$\langle N_{9EA} \rangle = \frac{[9EA]}{[Si\ NC]} = \frac{abs_{Si:9EA,395nm} - abs_{Si:dodecane,395nm} \frac{\langle abs_{Si:9EA} \rangle}{\langle abs_{Si:C12} \rangle}}{\varepsilon_{9MA,389nm}} \bigg/ \frac{abs_{Si:9EA,488nm}}{\varepsilon_{Si\ NC,488nm}} \quad (S1)$$

Here, $\langle abs_{Si:9EA} \rangle / \langle abs_{Si:C12} \rangle$ denotes to the normalization value used to match absorption spectra of Si:9EA and Si:dodecane across the 550 – 700 nm spectral range, $\varepsilon_{Si\ NC,488nm}$ is the molar extinction coefficient of the ~3.1 nm diameter Si QDs we employ at 488 nm (10,000 M$^{-1}$cm$^{-1}$),[4] and $\varepsilon_{9MA,389nm}$ corresponds to the molar extinction coefficient of 9-methylanthracene at 389 nm (8,413 M$^{-1}$cm$^{-1}$),[5] which corresponds to the maximum of its 0-0 vibronic transition. Note, in comparing the extinction spectra of 9MA to background subtracted spectra of Si:9EA, we account for a 6 nm red shift in the absorption maxima of the latter's 0-0 transition that results from the difference in the dielectric environments felt by 9MA molecules dissolved in toluene and 9EA molecules bound to the surface of Si QDs.[5]

A similar approach was used to determine $\langle N_{9VA} \rangle$, the number of anthracene molecules attached to the surface of Si:9VA, was performed using a measured molar extinction coefficient of 9VA of 6,900 M$^{-1}$cm$^{-1}$ at 389 nm. We note that due to strong coupling between Si QDs and 9VA molecules, absorption spectra of surface-bound 9VA molecules are significantly broadened relative that of 9VA molecules dispersed in solution (**Figure 1B**, **inset**). This will lead to a systematic underestimation of $\langle N_{9VA} \rangle$. However, as we are only interested in how the photoexcited dynamics and photon upconversion efficiency of Si:9VA change with increasing $\langle N_{9VA} \rangle$ and not the explicit dependence of these properties on a specific value of $\langle N_{9VA} \rangle$, this systematic underestimate does not impact any of the conclusions we draw in the main text.

| | | | | | | |
|---|---|---|---|---|---|---|
| Si:9EA | 9VA / μg | 8 | 31 | 125 | 500 | > 500 |
| | $\langle N_{9EA} \rangle$ | 1.29 | 2.98 | 5.38 | 6.72 | > 7 |
| Si:9VA | 9AA / μg | 8 | 31 | 125 | 500 | > 500 |
| | $\langle N_{9EA} \rangle$ | < 2 | 2.87 | 5.27 | 5.94 | 7.33 |

**Table S2:** The average number of attached 9EA or 9VA per Si QD, $\langle N_{9EA} \rangle$ or $\langle N_{9VA} \rangle$, with respect to different precursor loadings for the Si QD hydrosilylation process.

### 3.2. Photon Upconversion Quantum Efficiency Measurements

Photon upconversion measurements were performed using a 488 nm and 532 nm OBIS Coherent lasers as excitation sources and an Ocean Optics Maya 2000 Pro spectrometer for detection of upconverted light. Emitted photoluminescence (PL) was filtered through a ThorLabs notch filter before being received by the spectrometer to remove scattered excitation light. For each upconversion measurement, the system was calibrated using Rhodamine 6G (R6G) dissolved in ethanol as an emission reference (PLQY = 95%). The upconversion photon quantum yield (UCQY) was determined using the following expression:

$$\phi_{UC} = 2 \times \phi_{ref} \times \frac{photons\ absorbed\ by\ reference}{photons\ absorbed\ by\ upconversion\ solution} \times \frac{upconversion\ PL}{reference\ PL} \quad (S2a)$$

$$= 2 \times \phi_{R6G} \times \frac{1 - 10^{-OD_{R6G}}}{1 - 10^{-OD_{Si\ QD}}} \times \frac{n_{Emitter}^2}{n_{R6G}^2} \times \frac{[Area]_{Emitter}}{[Area]_{R6G}} \quad (S2b)$$

where $\phi_{ref}$ is the PLQY of R6G, $n_{Emitter}$ and $n_{R6G}$ are the refractive indices of the solvents for the upconversion, and R6G reference solutions (toluene and ethanol, respectively), $OD_{R6G}$ and $OD_{Si\ QD}$ denote the absorbance of Si QDs and R6G at the laser excitation wavelength, and $[Area]_{Emitter}$ and $[Area]_{R6G}$ are the integrated areas of the fluorescence peaks of the emitter (either DPA or *t*Bu$_4$P) and R6G, respectively. Note, this expression includes a factor of 2 to normalize the maximum UCQY that is achievable to a value of 100%.

We also note that the presence of surface-bound anthracene molecules are critical to obtaining photon upconversion. Upconversion measurements were performed using Si:dodecane particles that had been paired with 5.2 mM DPA emitters, but this resulted in an UCQY of only 0.18%. This was expected since dodecane presents a formidable tunneling barrier that hinders triplet energy transfer from Si QDs to DPA.

### 3.3. Correcting Photon Upconversion Efficiency Yields to Account for Reabsorption of Emitted Light

For photon upconversion measurements, if the emitter concentration is high, this can lead to a strong reabsorption of the upconverted light by emitter molecules in solution (inner filter effect), thereby lowering the upconversion efficiency. To assess how much this inner filter effect impacts measured upconversion quantum yield (UCQY) values, we have matched upconverted emission spectra with those measured by exciting the emitters directly in low concentration solutions wherein negligible reabsorption of emitted light occurs (**Figure S1**). Negligible reabsorption in these low concentration reference solutions was confirmed by verifying a lack of dependence of their emission lineshape on emitter concentration. Emission spectra measured from upconversion samples were normalized to those of the low-concentration references by matching spectra along their long-wavelength edges wherein emitters show no appreciable absorption. This allowed us to compute the ratio of the integrated area of the photon upconversion lineshape and the intrinsic emission lineshape of the emitter, which was then used to scale measured UCQYs to account for the reabsorption of emitted light. Taking into account this inner filter effect, UCQYs of Si:9EA:DPA and Si:9VA:*t*Bu$_4$P systems were computed to adopt maximum values of 15.8% and 17.2%, respectively, following optimization of the surface concentration of 9EA and 9VA.

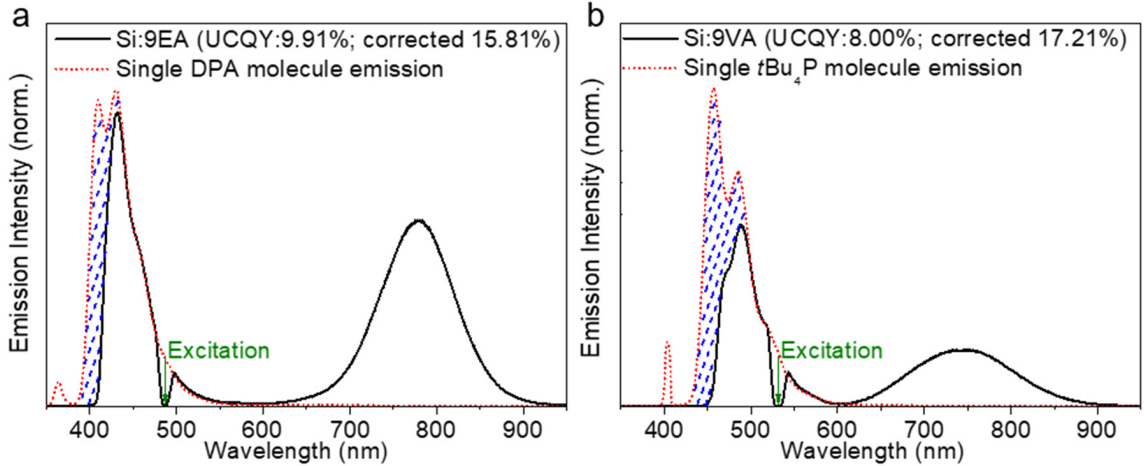

**Figure S1:** (a) PL spectra of Si:9EA:DPA ($\langle N_{9EA} \rangle$ = 2.98, black solid line, 5.2 mM DPA as emitter, excited at 488 nm) and low-concentration DPA (red dashed line, excited at 365 nm). (b) PL spectra of Si:9VA:$t$Bu$_4$P ($\langle N_{9VA} \rangle$ = 2.87, black solid line, 5.2 mM $t$Bu$_4$P as emitter, excited at 532 nm) and low-concentration $t$Bu$_4$P (2 µM, red dashed line, excited at 405 nm).

### 3.4. Dependence of Photon Upconversion Emission on Excitation Density

For photon upconversion measurements, the excitation density threshold is an important parameter that characterizes the performance of the upconversion system. Above the threshold, triplet fusion ceases to be rate-limiting as the steady-state population of emitters in excited spin-triplet states is sufficiently high that every excited emitter can undergo triplet fusion with an excited partner before their triplet excitations decay to the ground state.[6] Above this threshold, photon upconversion operates with its maximum quantum efficiency as only decay processes intrinsic to each participating species (Si QD, transmitter, and emitter) compete with upconversion. The point at which this threshold is reached is signaled by a change in the dependence of the intensity of upconverted light from a quadratic dependence on excitation density, and hence triplet concentration, to a linear dependence that is independent of triplet concentration.

**Figure S2a** plots the dependence of the upconverted emission produced by DPA when paired with Si:9VA on the intensity of a 488 nm excitation source while **Figure S2b** plots the dependence of upconverted emission generated by $t$Bu$_4$P when paired with Si:9VA on the intensity of a 532 nm excitation source. We find both systems exhibit low excitation density thresholds (1.5 W/cm$^2$ for Si:9EA and 0.5 W/cm$^2$ for Si:9VA). To the best of our knowledge, the excitation density threshold for Si:9VA is the lowest yet reported for Si QD-based upconversion systems.

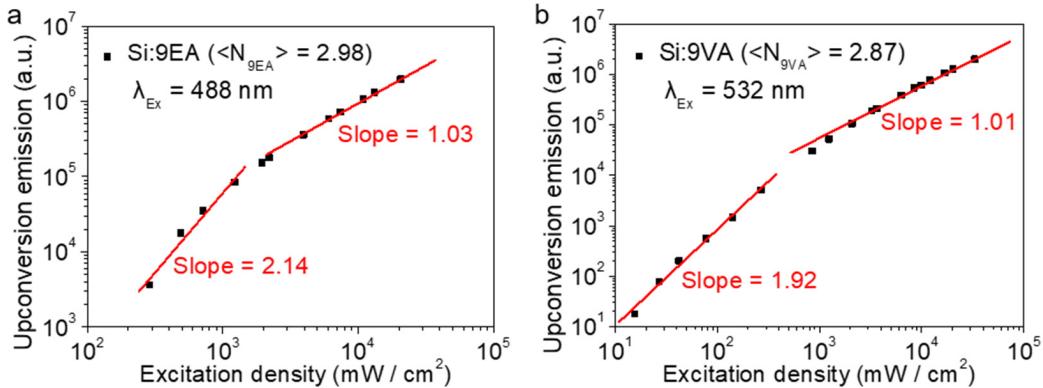

**Figure S2:** Log-log plots of the upconverted emission intensity of (a) Si:9EA ($\langle N_{9EA} \rangle$ = 2.98, DPA as emitter) versus laser excitation (488 nm) power and (b) the upconverted emission intensity of Si:9VA ($\langle N_{9VA} \rangle$ = 2.87, $t$Bu$_4$P as emitter) versus laser excitation (532 nm) power.

### 3.5. Triplet Sensitization of 9-vinylanthracene

To identify the triplet absorption spectrum of 9-vinyl anthracene (9VA) monomers, we performed a triplet sensitization experiment using platinum octaethylporphyrin (PtOEP). A mixture of 9VA and PtOEP in toluene was prepared and placed into a sealed cuvette. Photoexcitation of PtOEP's Q-band at 532 nm generates the lowest excited singlet state of PtOEP, which rapidly intersystem crosses to its triplet state on a ~165 fs timescale.[7] Since the energy of PtOEP's lowest triplet state is 1.9 eV,[8] which is higher than that of 9VA (~1.8 eV), diffusional collisions between photoexcited PtOEP and 9VA molecules can result in triplet energy transfer from PtOEP to 9VA, thus allowing us to identify 9VA's triplet absorption spectrum.

**Figure S3** plots TA spectra of solutions containing PtOEP and 9VA recorded following photoexcitation of PtOEP's Q-band. Two photobleachs appear near 385 nm and 505 nm that are respectively attributed to PtOEP's Soret and the first vibrational replica of PtOEP's Q-band. A broad photoinduced absorption centered at 420 nm is also seen that associated with PtOEP's $T_1$ state. As time evolves, features associated with PtOEP's $T_1$ state decay, concomitant with the appearance of a sharp resonance peaked at 435 nm, which we assign to the $T_1$ state of 9VA monomers. This peak agrees well with prior reports of triplet spectra of 9MA[5] and other anthracene derivatives.[9–11]

Importantly, the absorption spectrum of the $T_1$ state of monomeric 9VA differs substantially from that of 9VA molecules that have been attached to the surface of Si QDs (**Figure 4B**). We attribute this difference to strong coupling of 9VA's triplet exciton state to those of the Si QD, which gives rise to a band of states with mixed QD:molecule character.

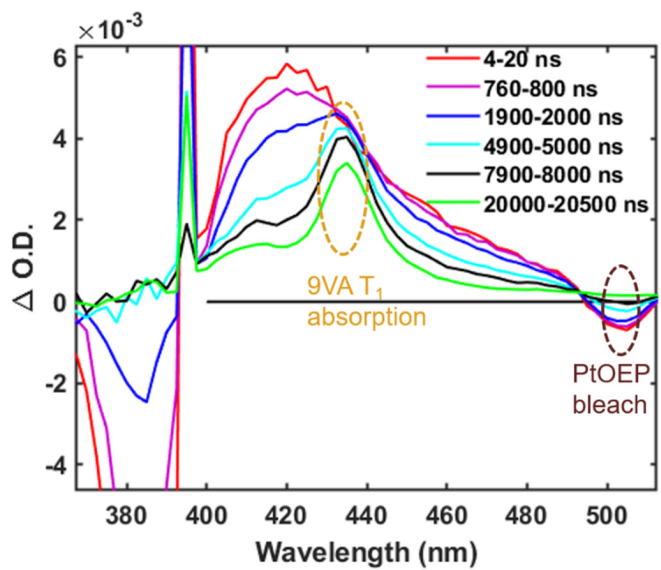

**Figure S3:** TA spectra of a PtOEP:9VA mixture excited at 532 nm.

### 3.6. Transient Absorption Spectra of Si:9EA as a Function of $\langle N_{9EA} \rangle$

To assess if strong electronic coupling exists between Si QDs and 9EA molecules bound to their surface, TA experiments were performed on Si:9EA as a function of $\langle N_{9EA} \rangle$. **Figure S4** plots TA spectra recorded 25 μs following excitation of Si:9EA at 532 nm. At this time delay, energy transfer from Si QDs to surface-bound 9EA molecules is fully complete, as evidenced by the appearance of a sharp resonance at 435 nm that corresponds to the $T_1 \rightarrow T_n$ absorption of surface-bound 9EA molecules.[5] Importantly, this band shows no change in its spectral position with increased loading of 9EA. This behavior is completely different from that shown by Si:9VA, which shows strong changes in its induced absorption spectra with changing 9VA surface concentration due to strong coupling between 9VA and silicon (**Figure 5B, main text**). The data in **Figure S4** indicates that the sp$^3$ C-C bridge that connects Si to anthracene in Si:9EA gives rise to only weak electronic coupling between these materials.

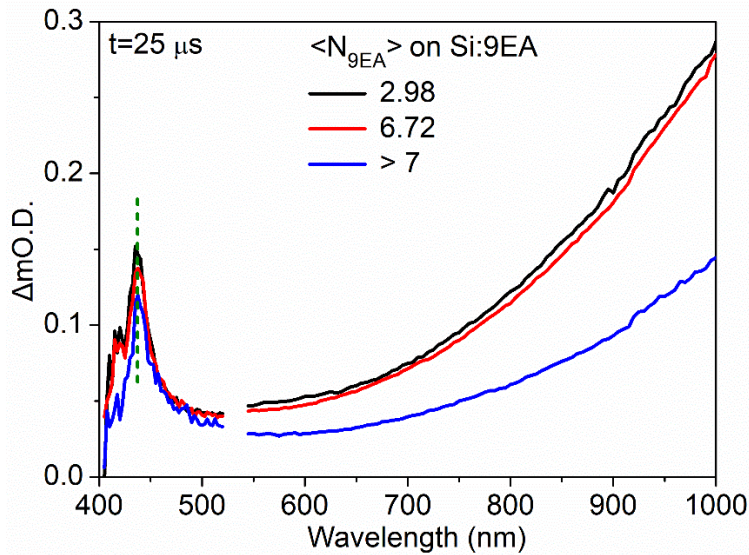

**Figure S4:** TA spectra of Si:9EA measured as a function of 9EA surface concentration ($\langle N_{9EA} \rangle$ = 2.98, 6.72, > 7). Spectra were recorded 25 µs following excitation at 532 nm.

### 3.7 Intersystem Crossing Timescale for Si:9VA

To determine the rate with which strongly-coupled triplet exciton states are populated following photoexcitation of Si:9VA, we employed a TA spectrometer with an instrument response function (IRF) of 1.6 ns. **Figure S5** plots the rise of the induced absorption signal of the strongly coupled triplet exciton state. A fit to this data reveals a growth timescale of 3 ns, which we attribute to intersystem crossing from bright Si QD state that is responsible for light absorption to a spatially-delocalized Si:9VA triplet exciton state.

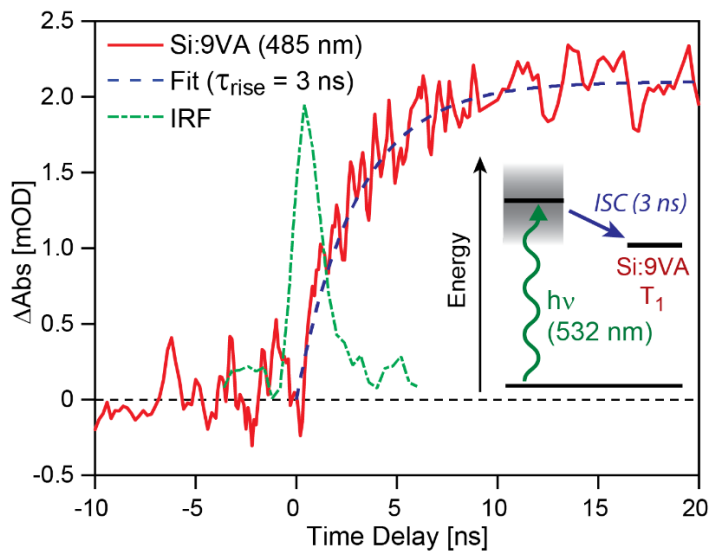

**Figure S5:** TA kinetics of Si:9VA (red solid) measured at a probe wavelength of 485 nm following photoexcitation of Si:9VA at 532 nm. At this probe wavelength we observe development of a spectrally broad induced absorption band (**Figure 2A**) that is attributed to a strongly-coupled triplet exciton state that spatially extends over both 9VA molecules and silicon. An exponential fit to the rise time of this band (blue dashed) yields a 3 ns time constant, that we attribute to the intersystem crossing timescale from the initial photoexcited Si QD state to the Si:9VA $T_1$ state (inset). This data was recorded using a nanosecond TA spectrometer with an instrument response function (IRF) of 1.6 ns (green dash dot).

## 4. Computational Methods

### 4.1. Silicon Model

For the computational modeling of the electronic structures of Si:9VA and Si:9EA, we employed a surface slab model consisting of six crystalline silicon layers terminated with Si(111) facets. Our slab model is semiperiodic: non-periodic in the surface normal-direction (z-axis), but periodically replicated along the plane that contains the terminating facets (x–y plane). Silicon's (111) facet was selected for computational modeling due to its stability relative to other silicon facets.[12–15] Si(111) also has each of its surface bonds oriented along the surface-normal direction, facilitating attachment of a single 9VA or 9EA molecule to the slab. All other silicon dangling bonds were passivated with hydrogen atoms. In the absence of any molecular attachments to silicon, this model yields two equivalent surface layers, each with 16 attachment sites, which corresponds to a 4×4 supercell expansion of the primitive Si(111) surface unit cell. This surface arrangement corresponds to a hexagonal cell, with equal in-plane lattice constants of 15.33 Å (**Figure S6**).

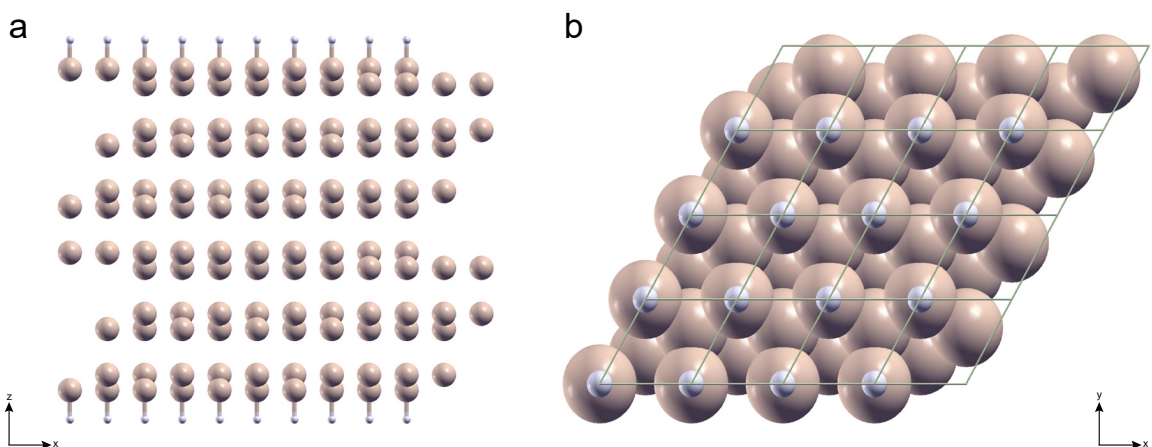

**Figure S6**: Surface slab supercell with no molecular attachments. (a) Side view of the surface slab supercell. Silicon atoms are tan and hydrogen atoms, which passivate the top-most layer of silicon, are gray. We do not show bonds between silicon atoms and only show the Si-H bonds at each surface so that the surface layers are distinct from the interior layers. Layers of silicon stacked in the z-direction are separated by whitespace, yielding 6 layers in total. (b) Top-down view of the surface slab supercell. We illustrate the atoms as space-filling spheres in this view so that the stacking of atoms below the surface layer is more obvious. The individual unit cell divisions are shown with small rhombuses, 16 in total, corresponding to a 4×4 supercell expansion, with the surface layer parallel to the x–y plane. The in-plane lattice constants are the same and equal to 15.33 Å. A molecular attachment can replace any hydrogen atom on either surface. We choose to attach 9VA or 9EA to the top-most surface when simulating the full Si:9VA or Si:9EA systems, respectively, thus generating an asymmetric slab model.

For a nonpolar hydrocarbon like 9VA or 9EA, we expect spurious interactions between molecules in periodic images to be small. We choose to put only a single 9VA or 9EA molecule on one of the slab's two surfaces—rather than one molecule on both surfaces—as the geometry of these molecules when attached to silicon is computationally expensive to converge. We leave the bottom 2 layers of silicon atoms frozen in their bulk positions during all geometry relaxations, and we relax all other atoms to a force convergence of less than 10 meV/Å.

We use density functional theory (DFT) computed with the Vienna Ab initio Simulation Package (VASP)[16–19] to interrogate the electronic structure of Si:9VA and Si:9EA. These calculations employ the projector augmented wave (PAW) method for the atomic potentials and the Perdew-Burke-Ernzerhof (PBE) generalized gradient approximation for the exchange–correlation functional.[20,21] Hydrocarbon molecules and silicon have similar electronegativities, and as a result, the dispersion interaction can be important.

However, bare DFT does not usually describe those interactions well. As such, we apply dispersion corrections to the PBE forces and energy using the DFT-D3(BJ) correction method of Grimme at al.[22,23] The PAW-PBE-D3(BJ) potential, in combination with a 540 eV plane-wave kinetic energy cutoff, yields a bulk silicon lattice constant of 5.42 Å, which is only a picometer smaller than experiment (5.43 Å).[24] To compute the triplet states in the slab calculations, we allow spin-polarization and force the calculation to find the lowest-energy triplet configuration, with a total spin component of $\pm\hbar$. These two spin components correspond to up and down spin channels, respectively.

### 4.2. Assessing Silicon:Molecule Coupling via the Density of States

The density of states (DOS) is the quantitative tool that we use to assess coupling between the electronic states of silicon and both 9VA and 9EA. The DOS is proportional to the trace of the imaginary part of the retarded Green's function:

$$G(E) = \frac{1}{E - H + i\varepsilon} \tag{S3}$$

where $\varepsilon$ is a small positive number that formally approaches zero from the right. The molecule-solid Hamiltonian is:

$$H = H_{mol} + H_{solid} + V \equiv H_0 + V, \tag{S4}$$

where $V$ is the coupling. When $V = 0$, the Green's function is a sum over poles that have resonances at the energies of the uncoupled molecule and solid. A nonzero coupling causes deviations from this form, and those deviations are related to the self-energy that characterizes the molecule-solid interaction.[25,26] In particular, for an eigenstate $|j\rangle$ of the molecular Hamiltonian, $H_{mol}$, one can define:

$$G_{jj}(E) = \langle j|\frac{1}{E - H + i\varepsilon}|j\rangle \tag{S4}$$

and the projected density of states (PDOS):

$$D_j(E) = -\frac{Im[G_{jj}(E)]}{\pi} = \sum_{\alpha=1}^{N} |\langle j|\alpha\rangle|^2 \delta(E - E_\alpha) \tag{S5}$$

where $H|\alpha\rangle = E_\alpha|\alpha\rangle$. When the coupling vanishes, $|\langle j|\alpha\rangle|^2 = \delta_{j=\alpha}$ and the PDOS has one and only one peak at the energy $\epsilon_j$ of the molecule. However, when there is coupling, this single resonance will shift and broaden—an effect quantified by the self-energy of the Fano-Anderson model,[25,26] but observable in the PDOS. Because it is prohibitively expensive to localize all relevant states of the molecule and the solid, as we have done in the past for the related problem of hole diffusion on the surfaces of semiconducting nanocrystals,[27–29] it is more convenient to analyze the PDOS than it is to compute the self-energy.

**Figure S7** highlights the DOS computed for Si:9VA and Si:9EA, which is summed over both spin up and spin down triplet channels. Note that due to our use of spin-unrestricted DFT, these DOS correspond to the effective single-particle orbitals for electrons and holes that, when paired, give rise to a triplet exciton configuration. The energy axis for both Si:9VA and Si:9EA is set such that the valence band edge corresponds to the zero of energy. States with energy below this value correspond to excited hole states within the valence band while those with energies above this value correspond to excited electron states within the conduction band. We compute the exciton DOS shown in **Figure 3A** of the main text by convolving the DOS for these two bands to give rise to triplet exciton configurations. We can show this by starting with the definition of the exciton DOS:

$$D_{ex}(E) = \sum_e \sum_h \delta(E - (\varepsilon_e - \varepsilon_h)) \tag{S6}$$

where $\varepsilon_e$ and $\varepsilon_h$ are the electron and hole energies, respectively. Using elementary properties of the delta

function, we find:

$$D_{ex}(E) = \int dE' \sum_e \delta(E' + E - \varepsilon_e) \sum_h \delta(E' - \varepsilon_h) \quad (S7)$$

or equivalently

$$D_{ex}(E) = \int dE' \rho_e(E' + E)\rho_h(E') \quad (S8)$$

where $\rho_{e/h}(E) = \sum_{e/h} \delta(E - \varepsilon_{e/h})$ is the DOS for the electrons/holes in the conduction/valence band.

**Figure S7** also shows the PDOS computed for the carbon p-orbitals of each system. We use these orbitals as surrogates for the frontier molecular orbitals (HOMO and LUMO) that would comprise the eigenstates of the isolated molecule. This is not a particularly accurate approximation, but the inaccuracies in making it are identical for 9VA as they are for 9EA, which allows us to infer differences in the coupling of these two materials to silicon by comparing their carbon p-orbital PDOS. As with the exciton density of states shown in **Figure 3A**, we calculate the exciton PDOS shown in that same figure by convolving the PDOS computed for the valence and conduction bands shown in **Figure S7**.

Examining the PDOS for Si:9EA, we find that these peaks are very narrow in energy, which is what we expect to see in the case wherein $V$ is small. This energy localization indicates minimal mixing between the molecular orbitals of 9EA with those of silicon. In contrast, the PDOS of Si:9VA show significant energetic broadening, indicating that the valence orbitals of 9VA hybridize with several silicon states. This gives rise to new, strongly-coupled triplet exciton states wherein charge carriers straddle the silicon:9VA interface, as visualized in **Figure 3C** of the main text.

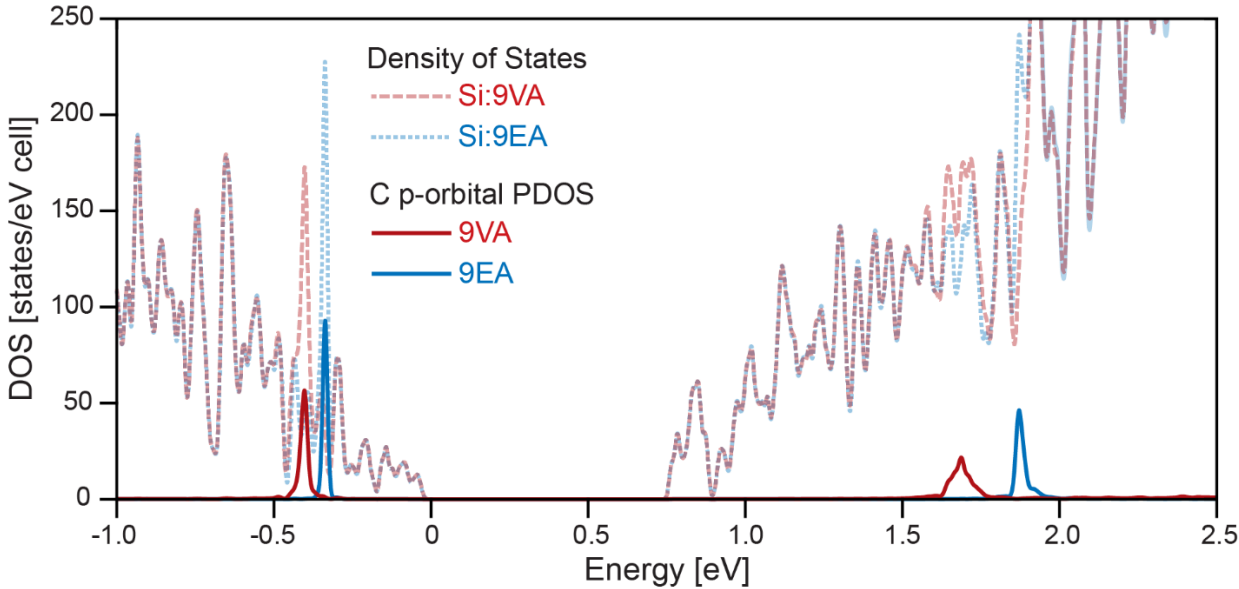

**Figure S7:** (A) Density of states (DOS) computed for Si:9VA (light red dashed) and Si:9EA (light blue dotted). Projected contributions to these states from the carbon p-orbitals of 9VA and 9EA are plotted as solid red and blue lines, respectively (carbon atom p-orbital, projected density of states: PDOS). These projections identify adiabatic states that contain contributions from the frontier orbitals (HOMO and LUMO) of these molecules. We set the valence band maximum as the zero of energy. The carbon p-orbital PDOS between -0.5 and 0 eV corresponds to a molecular HOMO peak, while the carbon p-orbital PDOS between 1.5 and 2 eV corresponds to a molecular LUMO peak. We note that the total DOS for each system is nearly identical, except in the vicinity of the HOMO and LUMO PDOS contributions to the total DOS.

### 4.3. Calculation of Partial Charge Densities

To aid in interpreting the computed DOS for Si:9VA and Si:9EA (**Figures 3A** & **S7**), we calculate the partial charge densities near the carbon p-orbital PDOS maxima to assess the electronic coupling between silicon and surface-bound anthracene molecules. These partial charge densities reveal the orbital density within a user-specified energy range, which allows us to visualize the degree to which these orbitals spatially extend over both the silicon lattice and anthracene molecules anchored to it.

**Figures 3B** & **3C** of the main text highlight the partial charge densities computed by examining a range of states whose energies fall within the peaks of the carbon p-orbital PDOS within the valence bands of Si:9VA and Si:9EA. We choose to focus on states in the valence band, which are constructed in part from the HOMOs of 9VA and 9EA, as they are occupied in the DFT calculation, which aids their computed accuracy. We note that the calculation of partial charge densities can be subject to user bias due to the need to define an energy range over which these densities are summed. For example, if molecules bound to the surface of silicon were completely uncoupled to it, one could obtain charge density plots that showed density spread across the silicon:molecule interface simply by choosing an energy integration range that was large enough to encompass the uncoupled states of both systems. Hence, our choice to examine states within the valence band rather than the conduction band of each system is also motivated by the lower total DOS near the peaks of the carbon p-orbital PDOS in the valence band. This makes a partial charge density analysis less prone to contamination by energetically close-lying silicon states that are not necessarily coupled to either 9VA or 9EA.

The partial charge density plots in **Figures 3B** & **3C** were constructed by summing the charge densities of states within the full-width at half-max (FWHM) of the carbon p-orbital PDOS peaks in the valence band for Si:9VA and Si:9EA. The FWHM corresponds to an energy range of 24.7 meV and 17.9 meV, respectively. These charge densities correspond to the spatial extent of the hole contribution to the triplet exciton states formed by 9VA and 9EA in each system. Examining these densities, we find that while the triplet exciton states remain spatially localized to 9EA, they spatially extend across the silicon|9VA interface, indicating the presence of weak electronic coupling in Si:9EA and strong electronic coupling in Si:9VA.

The use of a consistent integration range for both Si:9VA and Si:9EA should reduce the possibility that the difference in the charge density plots shown in **Figures 3B** & **3C** result not from the presence of strong coupling in Si:9VA, but rather from the inclusion of spurious uncoupled silicon states that happen to fall in the energy integration range. In choosing the FWHM for these two systems, we find that despite the difference in the energy ranges that generated the partial charge densities, the same number of eigenstates reside within each range: 20 eigenstates for each system in the first Brillouin zone, or 11 states in the irreducible Brillouin zone where the symmetry $\Psi_k = \Psi^*_{-k}$ is considered. We note that the former value across the full first Brillouin zone is more meaningful for this analysis; however, we mention both quantities since VASP only calculates the states within the irreducible Brillouin zone, along with their corresponding weights due to the aforementioned k-point symmetry.

We construct the FWHM partial charge density plots according to the wavefunctions of each system before calculating the DOS. We first calculate these wavefunctions self-consistently on a Γ-centered 3×3×1 k-point grid. Using the total charge density corresponding to these wavefunctions as inputs, we then calculate the subsequent DOS plots non-self-consistently on a Γ-centered 9×9×1 k-point grid. This allows us to define a FWHM of the relevant PDOS peaks in either system. The parent 3×3×1 k-point grid converges the total energy of the system to approximately 0.1 meV/atom, which ensures an accurate charge density. The 20 eigenstates within the FWHM of each system's carbon p-orbital PDOS in the valence band reside on the 3×3×1 k-point grid.